\documentclass[12pt]{article}
\usepackage{amssymb}
\usepackage{amsmath}
\usepackage{amstext}
\usepackage{graphicx,epsfig}
\usepackage{epsfig}
\usepackage{verbatim} 
\usepackage{fancyhdr}
\usepackage{fancybox}
\usepackage{color}
\usepackage{ulem,bbold}
\usepackage{enumitem}
\usepackage{subfigure}
\usepackage{bbm}
\usepackage{parskip}
\newcommand{\Comment}[1]{{}}
\definecolor{MyDarkBlue}{rgb}{0.15,0.15,0.45}
\usepackage[linktocpage=true]{hyperref}
\hypersetup{
colorlinks=true,
citecolor=MyDarkBlue,
linkcolor=MyDarkBlue,
urlcolor=MyDarkBlue,
}

\newcommand{\be}{\begin{equation}}
\newcommand{\ee}{\end{equation}}
\newcommand{\bea}{\begin{eqnarray}}
\newcommand{\eea}{\end{eqnarray}}
\newcommand{\beas}{\begin{eqnarray*}}
\newcommand{\eeas}{\end{eqnarray*}}
\newcommand{\nn}{\nonumber\\}

\def\({\left(}
\def\){\right)}

\newcommand{\bE}{\mathbf{E}}

\newcommand{\bF}{\mathbf{F}}

\newcommand{\bH}{\mathbf{H}}

\newcommand{\rd}{\mathrm{d}}

\numberwithin{equation}{section}

\begin{document}
\hspace{5.44in}IPMU14-0038

\begin{center}
{\Large \bf{Galileons Coupled to Massive Gravity: General Analysis and Cosmological Solutions}}
\end{center} 
 \vspace{.9truecm}
\thispagestyle{empty} \centerline{
{\large  { Garrett Goon ${}^{a,}$}}\footnote{E-mail address: \Comment{\href{mailto:: ggoon@physics.upenn.edu}}{\tt ggoon@physics.upenn.edu}} 
{\large  { A. Emir G\"umr\"uk\c{c}\"uo\u{g}lu ${}^{b,\, d,}$}}\footnote{E-mail address: \Comment{\href{mailto:Emir.Gumrukcuoglu@nottingham.ac.uk}}{\tt Emir.Gumrukcuoglu@nottingham.ac.uk}}
{\large  { Kurt Hinterbichler${}^{c,}$}}\footnote{E-mail address: \Comment{\href{mailto:khinterbichler@perimeterinstitute.ca}}{\tt khinterbichler@perimeterinstitute.ca}} 
}
\centerline{
{\large  { Shinji Mukohyama${}^{d,}$}}\footnote{E-mail address: \Comment{\href{mailto:shinji.mukohyama@ipmu.jp}}{\tt shinji.mukohyama@ipmu.jp}}
{\large  { Mark Trodden ${}^{a,}$}}\footnote{E-mail address: \Comment{\href{mailto:trodden@physics.upenn.edu}}{\tt trodden@physics.upenn.edu}}
                                                          }

\vspace{0.3 cm}

\centerline{{\it ${}^a$ 
Center for Particle Cosmology, Department of Physics and Astronomy,}}
 \centerline{{\it University of Pennsylvania, Philadelphia, PA 19104, USA}} 
 

\centerline{{\it ${}^b$ 
School of Mathematical Sciences, University of Nottingham,}}
 \centerline{{\it University Park, Nottingham, NG7 2RD, UK}} 
 

\centerline{\it $^c$ Perimeter Institute for Theoretical Physics,}
\centerline{\it 31 Caroline St. N, Waterloo, Ontario, Canada, N2L 2Y5}


\centerline{{\it ${}^d$ 
Kavli Institute for the Physics and Mathematics of the Universe, Todai Institutes for Advanced Study, }}
 \centerline{{\it University of Tokyo (WPI), 5-1-5 Kashiwanoha, Kashiwa, Chiba 277-8583, Japan}} 
 

\begin{abstract}
\vspace*{-.9em}
We further develop the framework for coupling galileons and Dirac-Born-Infeld (DBI) scalar fields to a massive graviton while retaining both the non-linear symmetries of the scalars and ghost-freedom of the theory.  The general construction is recast in terms of vielbeins which simplifies calculations and allows for compact expressions.  
Expressions for the general form of the action are derived, with special emphasis on those models which descend from maximally symmetric spaces.  We demonstrate the existence of maximally symmetric solutions to the fully non-linear theory and analyze their spectrum of quadratic fluctuations. Finally, we consider self-accelerating cosmological solutions and study their perturbations, showing that the vector and scalar modes have vanishing kinetic terms.
\end{abstract}

\newpage

\thispagestyle{empty}
\newpage
\setcounter{page}{1}
\setcounter{footnote}{0}

\section{Introduction}\label{Introduction}
\parskip=5pt
\normalsize

Various modifications of gravity have been proposed in order to explain the observed cosmic acceleration, among other reasons. 
The study of certain classes of ghost-free models has led to interest in scalar fields referred to as ``galileons" which enjoy non-linear symmetries of the form
\begin{align}
\pi(x)&\to \pi(x)+c+b_{\mu}x^{\mu}\ ,\label{introgalileansymmetry}
\end{align}
where $\pi(x)$ is a scalar field and $c$ and $b_{\mu}$ are constant.
All such Lagrangians which exhibit the above symmetry and whose equations of motion remain second order have been classified and generalized \cite{Nicolis:2008in} (see \cite{Trodden:2011xh,deRham:2012az,Deffayet:2013lga} for reviews).
These theories have proven interesting for both phenomenological and theoretical reasons.  On the phenomenology side, galileon theories exhibit the Vainshtein screening mechanism \cite{Vainshtein:1972sx} (see \cite{Clifton:2011jh,Babichev:2013usa,Khoury:2013tda} for reviews) which can potentially keep them in accord with current fifth-force experimental bounds through the effects of large classical gradients.  Furthermore, there exists evidence that galileons are well-behaved quantum mechanically due to a non-renormalization theorem which states that galileons are not corrected by self-interaction loops \cite{Luty:2003vm,Nicolis:2004qq,Hinterbichler:2010xn,deRham:2012ew}. Importantly, this ensures that quantum corrections are irrelevant and classical calculations can be trusted in the Vainshtein screening regime where gradients of $\pi$ are large.  On the theoretical side, a geometric viewpoint in which galileons arise due to the presence of 4D brane in a 5D bulk was developed in \cite{deRham:2010eu} and generalized in \cite{Goon:2011uw,Goon:2011qf,Burrage:2011bt,Goon:2011xf}.  The galileons are interpreted as the Goldstone modes corresponding to the spontaneous breaking of  spacetime symmetries due to the presence of the brane in the bulk.  Using canonical methods for analyzing spontaneous symmetry breaking, it can be shown that the galileon Lagrangians correspond to Wess-Zumino terms for the appropriate symmetry breaking pattern \cite{Goon:2012dy}.

A satisfactory method of coupling galileon theories to gravity while retaining their desirable properties has proven elusive, however.  Minimal coupling of galileons to gravity leads to equations of motion which have higher order derivatives of the metric.  Non-minimal couplings can be added to yield second order equations of motion, but this alteration breaks the galileon symmetries \cite{Deffayet:2009mn,Deffayet:2009wt}. 

In \cite{Gabadadze:2012tr} a procedure was developed for coupling galileons (and DBI scalars, more generally) to a dynamical metric, $g_{\mu\nu}$, while retaining all of the desired properties of the theory.  In this framework, $g_{\mu\nu}$ describes a \textit{massive} graviton.  The fully non-linear theory of a massive graviton was only recently elucidated \cite{deRham:2010ik,deRham:2010kj} (see \cite{Hinterbichler:2011tt,deRham:2014zqa} for reviews) and it is this de Rham-Gabadadze-Tolley (dRGT) theory which most naturally incorporates the galileon.  The theory of \cite{Gabadadze:2012tr} non-linearly propagates the correct number of degrees of freedom for a scalar coupled to a massive graviton, with no Boulware-Deser ghost mode~\cite{Andrews:2013ora}, and the galileon symmetry remains intact.

Here, we further study this theory of galileons and DBI scalars coupled to a metric.  In Section \ref{GalileonConstruction} we briefly review the probe brane derivation of generic galileon theories, the dRGT theory of massive gravity, and the coupling of galileons to massive gravity.  
In Section \ref{generalconstruction} we derive some new formulae for arbitrary bulk metrics, including the the cases of maximally symmetric bulk metrics, which result in the greatest number of non-linear galileon symmetries.  In Section \ref{maximallysymmetricsolutions} we find maximally symmetric solutions to the full non-linear theory and study the spectrum of fluctuations about them.  Finally, in Section \ref{selfacceleratingsolutionsandperturbations} we discuss self-accelerating cosmological solutions and explore their perturbations.

\textbf{Conventions:} The mostly plus signature is used and we follow \cite{Hinterbichler:2012cn} for all tensor conventions.  In particular, we choose the flat, Levi-Civita \textit{symbol} to follow the convention $\tilde{\epsilon}_{01\ldots d}=+1$ as well as $\tilde{\epsilon}^{01\ldots d}=+1$ so that
\begin{align}
\tilde{\epsilon}^{c_{1}\ldots c_{p}a_{1}\ldots a_{d-p}}\tilde{\epsilon}_{c_{1}\ldots c_{p}b_{1}\ldots b_{d-p}}&=p!(d-p)!\delta^{[a_{1}}_{b_{1}}\ldots\delta^{a_{d-p}]}_{b_{d-p}}
\end{align}
and $\tilde{\epsilon}^{a_{0}\ldots a_{d}}=-\eta^{a_{0}b_{0}}\ldots\eta^{a_{d}b_{d}}\tilde{\epsilon}_{b_{0} \ldots b_{d}}$.
We symmetrize and anti-symmetrize tensors with weight 1 so that, for example, $M_{[ab]}=(M_{ab}-M_{ba})/2$ and $M_{(ab)}=(M_{ab}+M_{ab})/2$.  

\section{Review of Galileon Brane Construction and Massive Gravity}\label{GalileonConstruction}

In this section we briefly review the ingredients and construction of the Galileon and massive gravity theories we are interested in.

\subsection{Galileon Probe-Brane Construction\label{probebranesubsection}}

We start with a brief review of the probe brane construction of galileons and DBI scalars on general curved backgrounds.  For more details see \cite{Goon:2011qf,Goon:2011uw,Gabadadze:2012tr}.  While the following can be generalized to the case of multiple galileons along the lines of \cite{Hinterbichler:2010xn}, we restrict ourselves to the single galileon case\footnote{One can also consider bulk spaces with two temporal directions.  We do not explicitly consider this case here, but the results are relevant to existing theories such as the extension of quasi-dilation massive gravity in \cite{DeFelice:2013tsa}.}.  One begins by considering a 4+1 dimensional bulk with coordinates $X^{A}$, a fixed bulk metric $G_{AB}(X)$, 
and an embedded 3-brane with world-volume coordinates $x^{\mu}$.  The brane position is given by the embedding functions $X^{A}(x^{})$.  The embedding functions  define a set of four 5D tangent vectors $\frac{\partial X^{A}}{\partial x^{\mu}}\partial_{A}\equiv e^{A}{}_{\mu}\partial_{A}$ and a normal vector $n^{A}$ satisfying
\begin{align}
0=G_{AB}n^{A}e^{B}{}_{\mu},\ \ \ 
1=G_{AB}n^{A}n^{B}\ ,
\end{align}
which in turn define the 4D extrinsic curvature tensor,
\begin{align}
K_{\mu\nu}&=-n_{A}e^{B}{}_{\mu}\nabla_{B}e^{A}{}_{\nu}\ .
\end{align}

We wish to build actions on the brane and we demand that they be invariant under brane diffeomorphisms, $x^{\mu}\to x^{\mu}-\xi^{\mu}$. The only covariant ingredients at our disposal are then the induced metric 
\be \bar{g}_{\mu\nu}=\partial_{\mu}X^{A}\partial_{\nu}X^{B}G_{AB}(X),\label{inducedmetricform}\ee
the covariant derivative compatible with the induced metric $\bar{\nabla}_{\mu}$, its curvature $\bar{R}_{\mu\nu\rho\sigma}$, and the extrinsic curvature $K_{\mu\nu}$, 
\begin{align}
S_{\rm }&=\int \rd^{4}x\, \sqrt{-\bar g}\, \mathcal{L}_{\rm }\left (\bar{g}_{\mu\nu},\bar{R}_{\mu\nu\rho\sigma},K_{\mu\nu},\bar{\nabla}_{\mu}\right ) \ .\label{galileoningredients}
\end{align}
The dynamical variables are the five embedding functions $X^{A}$.  Brane diffeomorphism invariance will render four of these unphysical, leaving a single physical brane-bending degree of freedom.  

It is convenient to work in a fixed gauge where this single degree of freedom is made manifest and the most natural choice is ``unitary gauge" (or static gauge) in which the first four embedding functions are chosen to coincide with the brane coordinates and the fifth becomes the galileon field, $\pi$,
\begin{align}
X^{\mu}=x^{\mu},\ \ \ X^{5}&=\pi(x^{})
\ .\label{galileonembeddingfunctions}
\end{align}
$\pi(x)$ then measures the fluctuations of the brane transverse to some hypersurface $X^{5}=constant$.

The symmetries of the theory are inherited from bulk Killing vectors; for each bulk killing vector $K^{A}(X)$ the transformation 
\be \delta X^{A}=K^{A}(X)\label{globalngfsym}\ee
 is a global symmetry.  If we have fixed a gauge, then this transformation may ruin our gauge choice and we must re-fix the gauge by a compensating brane diffeomorphism.  In the case of unitary gauge \eqref{galileonembeddingfunctions}, the global symmetry acts as $x^{\mu}\to x^{\mu}+K^{\mu}(\pi,x^{})$, $\pi\to \pi+K^{5}(\pi,x^{})$,
so to re-fix the gauge we must perform a brane diffeomorphism with $\xi^{\mu}=K^{\mu}(\pi,x^{})$, so that the total, gauge-preserving global symmetry is given by \cite{Goon:2011qf}
\begin{align}
x^{\mu}&\to x^{\mu}\nn
\pi&\to \pi+K^{5}(\pi,x^{})-K^{\mu}(\pi,x^{})\partial_{\mu}\pi\ .\label{generalgaugepreservinggalileonsymmetries}
\end{align}
This is the more general case of the galileon symmetry \eqref{introgalileansymmetry}.

The final defining characteristic of galileon theories is that the equations of motion remain second order, despite the higher derivatives appearing in the action.  This condition will not be satisfied for a generic choice of action.  It is only satisfied when the action is comprised of the 4D Lovelock curvature invariants and the Gibbons-Hawking-York boundary terms associated with 5D Lovelock invariants \cite{deRham:2010eu}.  These terms are
\begin{align}
\mathcal{L}_{2}&=  -\sqrt{-\bar{g}}  \nn
\mathcal{L}_{3}&=  \sqrt{-\bar{g}}K  \nn
\mathcal{L}_{4}&=  -\sqrt{-\bar{g}}\bar{R}  \nn
\mathcal{L}_{5}&=  \frac{3}{2}\sqrt{-\bar{g}}\Big [-\frac{1}{3}K^{3}+K_{\mu\nu}^{2}K-\frac{2}{3}K_{\mu\nu}^{3}  -2\left (\bar{R}_{\mu\nu}-\frac{1}{2}\bar{R}\bar g_{\mu\nu}\right )K^{\mu\nu}\Big ]\ ,\label{lovelock}
\end{align}
in addition to a non-derivative tadpole term which, in unitary gauge, takes the form~\cite{Goon:2011qf}
\begin{align}
\mathcal{L}_{1}&= \int\rd^{4}x\, \int^{\pi(x)}\rd\pi'\sqrt{|G_{AB}(x^{},\pi')|}.
\label{galileontadpoleterm}
\end{align}
The phrase ``galileon action" refers to the sum of these special terms, 
\be S_{\rm gal}=\int\rd^{4}x\,\sqrt{-\bar g}\mathcal{L}_{\rm gal}=\sum_{i=1}^{5}\int\rd^{4}x\,c_{i}\mathcal{L}_{i},\label{galactionsep}\ee
with the $\mathcal{L}_{i}$ as defined in \eqref{lovelock} and \eqref{galileontadpoleterm}.  Generic theories constructed in this manner are alternatively referred to as galileon type theories or DBI-galileon theories (${\cal L}_2$ is the traditional DBI term).

\subsection{Ghost-free Massive Gravity and Interacting Spin-2 Fields}\label{MassiveGravityandInteractingSpin2s}

A challenge one encounters when attempting to develop an interacting theory of a massive graviton by adding a potential to the Einstein-Hilbert term is the generic presence of a sixth degree of freedom, the Boulware-Deser ghost \cite{Boulware:1973my}.   The dRGT theory \cite{deRham:2010ik,deRham:2010kj} tunes the potential in such a manner as to remove the offending degree of freedom \cite{Hassan:2011ea,Hassan:2011hr,deRham:2011rn,deRham:2011qq,Mirbabayi:2011aa, Golovnev:2011aa,Hassan:2012qv,Hinterbichler:2012cn,Kluson:2012wf}.  The dRGT action takes the form
\begin{align}
S_{\rm dRGT}&=\frac{M_{\rm pl}^{2}}{2}\int \rd ^{4}x\, \sqrt{-g}\left [R[g]-2\Lambda-\frac{m^{2}}{4}\sum_{n=1}^{3}\beta_{n}S_{n}\left (\sqrt{g^{-1}\eta}\right )\right ] \ , 
\label{dRGTactionsquarerootform}
\end{align}
where $g_{\mu\nu}$ is the dynamical metric, $\eta_{\mu\nu}$ is a fixed Minkowski fiducial metric, and $S_{n}$ is the $n$-th elementary symmetric polynomial of the matrix square root of $g^{\mu\rho}\eta_{\rho\nu}$, given by
\begin{align}
S_{n}\left (M^{\mu}{}_{\nu}\right )&\equiv \frac{1}{n!(4-n)!}\tilde\epsilon_{\mu_{1}\ldots \mu_{4}}\tilde\epsilon^{\nu_{1}\ldots \nu_{4}}M^{\mu_{1}}{}_{\nu_{1}}\ldots M^{\mu_{n}}{}_{\nu_{n}}\delta^{\mu_{n+1}}_{\nu_{n+1}}\ldots \delta^{\mu_{4}}_{\nu_{4}}
\end{align}
for a $4\times 4$ matrix $M^{\mu}{}_{\nu}$, and $\tilde\epsilon$ is the flat space Levi-Civita symbol (the $n=0$ symmetric polynomial is omitted from \eqref{dRGTactionsquarerootform} since it is degenerate with the cosmological constant $\Lambda$, and $n=4$ is omitted because it is a constant).  The dRGT theory can be extended to a theory of two interacting metrics by promoting the fixed $\eta_{\mu\nu}$ in \eqref{dRGTactionsquarerootform} to a dynamical metric $f_{\mu\nu}$ and adding an Einstein-Hilbert term and cosmological constant for $f_{\mu\nu}$.  The resulting bi-gravity theory is also free of the Boulware-Deser ghost \cite{Hassan:2011zd} and describes a massless graviton interacting with a massive one.

Since it can be unwieldy to work with matrix square roots, we will primarily make use of an equivalent\footnote{See, however, \cite{Deffayet:2012zc} for some caveats.} dRGT construction in terms of vielbeins.  After writing the metric in terms of vielbeins\footnote{We write vielbein 1-forms in bold such as $\bE^{a}$ and label their components as $E_{\mu}{}^{a}$, i.e. $\bE^{a}=E_{\mu}{}^{a} \rd x^{\mu}$.}, $g_{\mu\nu}=E_{\mu}^{\ a} E_\nu^{\ b}\eta_{ab}$, $a\in\{0,1,2,3\}$ and $\eta_{ab}=diag(-,+,+,+)$, and introducing the unit one-form $\mathbf{1}^{a}=\delta^{a}_{\mu}\rd x^{\mu}$, the symmetric polynomials of $\sqrt{g^{-1}\eta}$ can be written as
\begin{align}
\rd^{4}x\, \sqrt{-g}\,S_{n}\left (\sqrt{g^{-1}\eta}\right )&\propto  \tilde{\epsilon}_{a_{1}\ldots a_{4}}\mathbf{1}^{a_{1}}\wedge\ldots\wedge\mathbf{1}^{a_{n}}\wedge\bE^{a_{n+1}}\wedge\ldots\wedge\bE^{a_{4}} \ ,
\end{align}
so that the dRGT action is rephrased as
\begin{align}
S_{\rm dRGT}&=\frac{M_{\rm pl}^{2}}{2}\Big[\int \rd ^{4}x\, \left (\det E_{}{}^{}\right )\, R[E_{}{}^{}]-2\Lambda\nn
&\quad -\frac{m^{2}}{4}\sum_{n=1}^{3}\frac{\beta_{n}}{n!(4-n)!}\int\, \tilde{\epsilon}_{a_{1}\ldots a_{4}}\mathbf{1}^{a_{1}}\wedge\ldots\wedge\mathbf{1}^{a_{n}}\wedge\bE^{a_{n+1}}\wedge\ldots\wedge\bE^{a_{4}}\Big ]\ .
\end{align}
The six extra components present in the vierbein which are not present in the metric are eliminated algebraically by their own equations of motion, see \cite{Hinterbichler:2012cn} for details.  It can be more technically efficient and conceptually clearer to work with the vielbein variables.  For instance, the Hamiltonian constraint analysis is more straightforward in the vielbein language and vielbeins are the natural variables with which one describes more general theories of multiple interacting spin-2 degrees of freedom \cite{Hinterbichler:2012cn}.

\subsection{Coupling Galileons and DBI Scalars to a Metric}\label{couplinggalileonsanddbiscalarstoametric}

We now review the construction of  \cite{Gabadadze:2012tr}, which incorporates dRGT massive gravity into the braneworld construction of galileons.  The basic ingredients at our disposal are the induced brane metric \eqref{inducedmetricform} of Section \ref{probebranesubsection}, $\bar{g}_{\mu\nu}$, which contains the galileon or DBI degrees of freedom, and the dynamical metric, $g_{\mu\nu}$, which lives on the brane. We couple these together by writing the dRGT action \eqref{dRGTactionsquarerootform} and replacing the flat metric $\eta_{\mu\nu}$ by the induced metric $\bar{g}_{\mu\nu}$,
\begin{align}
  S&=\frac{M_{\rm pl}^{2}}{2}\int \rd ^{4}x\, \sqrt{-g}\left [R[g]-2\Lambda-\frac{m^{2}}{4}\sum_{n=1}^{3}\beta_{n}S_{n}\left (\sqrt{g^{-1}\bar{g}}\right )\right ]+S_{\rm gal}[\bar{g}]\ .\label{matrixformofgalileongravitonaction}
  \end{align}  
We have also added the action $S_{\rm gal}$ in \eqref{galactionsep}, comprised of the galileon Lagrangians \eqref{lovelock}, which gives further dynamics to the galileon sector but does not introduce additional couplings between $g_{\mu\nu}$ and $\bar g_{\mu\nu}$.
 
 Non-linear symmetries of the fixed bulk metric \eqref{globalngfsym} continue to be symmetries of \eqref{matrixformofgalileongravitonaction} despite the dynamical metric.  Once unitary gauge \eqref{galileonembeddingfunctions} is fixed, these symmetries will act on the metric via the compensating brane diffeomorphism, as described in \cite{Gabadadze:2012tr}.  The whole construction remains free of the Boulware-Deser ghost, so there are six degrees of freedom non-linearly: five for the massive graviton and one for the galileon \cite{Andrews:2013ora}.
 
We may rephrase the above theory in the vielbein formalism \cite{Andrews:2013ora}.  We write both the physical metric and the induced metric in terms of vielbeins  
\be g_{\mu\nu}=E_\mu^{\ a} E_\nu^{\ b}\eta_{ab}, \ \ \ \bar g_{\mu\nu}={\bar E}_\mu^{\ a} {\bar E}_\nu^{\ b}\eta_{ab}.\ee
For the induced metric, we will choose the vierbein to be in upper triangular form 
 \be {\bar E}_{\mu}{}^{a}=\begin{pmatrix}
 {\bar N}& {\bar N}^{i}\bar e_{i}{}^{\hat a}\\ 0 & \bar e_{i}{}^{\hat a} \label{f:vierbeins}
\end{pmatrix} \ .
\ee
where $i,j,\ldots $ are spatial coordinate indices raised and lowered with the spatial metric $\bar g_{ij}$, and $\hat a,\hat b,\ldots$ are spatial Lorentz indices raised and lowered with $\delta_{\hat a\hat b}$.
Here $\bar N$ and $\bar N^i$ are ADM \cite{Arnowitt:1960es} lapse and shift variables, and $\bar e_{i}{}^{\hat a}$ is an upper triangular spatial dreibein for the spatial part of the induced metric and $\bar e^{i}{}_{\hat a}$ its inverse transpose.  These are obtained in terms of the embedding field $X^{ A}$ by solving
\begin{align}
{\bar g}_{00}=\dot{X}^{ A}\dot{X}^{ B}G_{{ A}{ B}}(X)&=- {\bar N}^{2}+ {\bar N}^{i} {\bar N}_{i} \nonumber\\
{\bar g}_{0i}=\dot{X}^{ A}\partial_{i}X^{ B}G_{{ A}{ B}}(X)&= {\bar N}_{i} \nonumber\\
{\bar g}_{ij}=\partial_{i}X^{ A}\partial_{j}X^{ B}G_{{ A}{ B}}(X)&=\bar e_{i}{}^{a}\bar e_{j}{}^{b}\delta_{ab}\ .\label{lsforphi}
\end{align}
The upper triangular vierbein \eqref{f:vierbeins} has 10 components, and is just a re-packaging of the 10 components in $\bar g_{\mu\nu}$, which in turn depend only on the $X^{ A}$.
We may now use the interacting vielbein formalism of \cite{Hinterbichler:2012cn} to construct a vierbein action equivalent to \eqref{matrixformofgalileongravitonaction}, 
\begin{align}
S&=\frac{M_{\rm pl}^{2}}{2}\Big [\int \rd ^{4}x\, \left (\det E\right )\, \left [R[E]-2\Lambda\right ]\nn
&\quad-\frac{m^{2}}{4}\sum_{n=1}^{3}\frac{\beta_{n}}{n!(4-n)!}\int\, \tilde\epsilon_{a_{1}\ldots a_{4}}\bar\bE^{a_{1}}\wedge\ldots\wedge\bar\bE^{a_{n}}\wedge\bE^{a_{n+1}}\wedge \ldots\bE^{a_{4}}\Big ]+S_{\rm gal}[\bar{E}]\label{galileonmetricmixingaction}\ .
\end{align}
As in the pure massive gravity case, the six extra components present in the dynamical vierbein $E_\mu^{\ a}$ which are not present in the dynamical metric $g_{\mu\nu}$ are eliminated algebraically by their own equations of motion.

\section{General Construction}\label{generalconstruction}

Much of the remainder of this paper is devoted to using the vielbein formalism to derive some explicit expressions for the action in various limiting cases which are more general than those studied in~\cite{Gabadadze:2012tr}.
In what follows, we place special emphasis on cases where the bulk metric is maximally symmetric.  

The interesting terms in the action \eqref{galileonmetricmixingaction} are those which mix the $\bE^{a}$ and $\bar{\bE}^{a}$ vielbeins and we define the ``mixing action" to be
\begin{align}
S_{\rm mixing}&=-\frac{1}{8}M_{\rm pl}^{2}m^{2}\sum_{n=1}^{3}\frac{\beta_{n}}{n!(4-n)!}S^{(n)}_{\rm mix},\nn 
S^{(1)}_{\rm mix}&\equiv\int \, \tilde{\epsilon}_{abcd}\, \bar\bE^{a}\wedge\bE^{b}\wedge\bE^{c}\wedge\bE^{d} , \nn
S^{(2)}_{\rm mix}&\equiv\int \, \tilde{\epsilon}_{abcd}\, \bar\bE^{a}\wedge\bar\bE^{b}\wedge\bE^{c}\wedge\bE^{d} , \nn
S^{(3)}_{\rm mix}&\equiv\int \, \tilde{\epsilon}_{abcd}\, \bar\bE^{a}\wedge\bar\bE^{b}\wedge\bar\bE^{c}\wedge\bE^{d} .\label{mixingactions}
\end{align}

\subsection{Gaussian Normal Form}

It proves convenient to express the 5D metric in Gaussian normal form
\begin{align}
G&=\rd\rho^{2}+f_{\mu\nu}(X^{\sigma},\rho)\rd X^{\mu}\rd X^{\nu},\label{gaussiannormal5Dmetric}
\end{align}
where we've labeled the 5D coordinates as $\rho=X^{5}$ and $X^{\mu}$, $\mu\in\{0,1,2,3\}$.  We introduce a vielbein $\bF^{i}$ on the 5D space $\mathcal{N}$,
\be G=G_{AB}\rd X^{A}\rd X^{B}=\bF^{i}\otimes\bF^{j}\eta_{ij}=F_{A}{}^{i}F_{B}{}^{j}\eta_{ij}\rd X^{A}\rd X^{B}\ ,\label{5Dgeneralconstructionmetricandvielbeins} \ee
where $i\in\{0,1,2,3,5\}$ and $\eta_{ij}=diag(-,+,+,+,+)$.
The Gaussian normal vielbein components satisfy
\begin{align}
F_{5}{}^{i}F_{5}{}^{j}\eta_{ij}&=1\nn
F_{5}{}^{i}F_{\mu }{}^{j}\eta_{ij}&=0\nn
F_{\mu}{}^{i}F_{\nu}{}^{j}\eta_{ij}&=f_{\mu\nu}(X^{\sigma},\rho)\label{gaussiannormalvielbeins}
\end{align}
and we will take $F_{5}{}^{i}=\delta^{i}_{5}$ and $F_{\mu}{}^{5}=0$ with the remaining components $F_{\mu}{}^{i}$ determined by taking some solution to the last equation in \eqref{gaussiannormalvielbeins}.

Relabeling the embedding functions as $X^{5}\equiv\pi$ and $X^{\mu}$, $\mu\in\{0,1,2,3\}$, the pullback of the 5D metric becomes
\begin{align}
(X^{*}G)&=\bar{g}\nn
&=\left (f_{\alpha\beta}\left (X^{\sigma},\pi\right )\partial_{\mu}X^{\alpha}\partial_{\nu}X^{\beta}+\partial_{\mu}\pi\partial_{\nu}\pi\right )\rd x^{\mu}\rd x^{\nu}\nn
&=F_{\alpha}{}^{a}F_{\beta}{}^{b}\eta_{ab}\rd X^{\alpha}\rd X^{\beta}+\rd\pi\rd\pi\nn
&\equiv \bar{\bE}^{a}\bar{\bE}^{b}\eta_{ab}\ .\label{pullbackvielbeins}
\end{align}
Solving for $\bar E_{\mu}{}^{a}$ yields
\begin{align}
\bar{E}_{\mu}{}^{a}&=F_{\nu}{}^{a}\left (X^{\sigma},\pi\right )\partial_{\mu}X^{\nu}+\kappa\Pi^{a}\partial_{\mu}\pi,\nn
\end{align}
where
\begin{align}
\Pi^{a}&\equiv \eta^{ab}F^{\nu}{}_{b}\frac{\partial x^{\mu }}{\partial X^{\nu}}\partial_{\mu}\pi\nn
\kappa&\equiv \frac{1}{\Pi^{2}}\left [-1+\sqrt{1+\Pi^{2}}\right ],\quad \Pi^{2}\equiv \eta_{ab}\Pi^{a}\Pi^{b}\ ,\label{generalvielbeinform}
\end{align}
and we have assumed that $\frac{\partial X^{\nu}}{\partial x^{\mu}}$ is invertible, with inverse $\frac{\partial x^{\mu}}{\partial X^{\nu}}$.  The sign of the square root in $\kappa$ is taken to be positive so that $\kappa$ is analytic as $\Pi^{a}\to0$.

 \subsection{Mixing Actions in Component Form}
 
 We now present some expressions for the mixing actions \eqref{mixingactions}, in terms of $F_{\mu}{}^{a}$,  $E_{\mu}{}^{a}$, $X^{\sigma}$ and $\pi$.  All cases can be expressed in terms of flat space Levi-Civita symbols as follows\footnote{Here we use the following general expression for the wedge products of two sets of vielbeins, $\omega^{A}$ and $\Omega^{A}$, in $D$-dimensions,
\begin{align}
&\quad\tilde{\epsilon}_{A_{1}\ldots A_{D}}\omega^{A_{1}}\wedge\ldots\wedge\omega^{A_{d}}\wedge\Omega^{A_{d+1}}\wedge\ldots\wedge\Omega^{A_{D}},\nn
&=\det\left ( \omega_{\mu}{}^{C}\right )\rd^{D}x\, \tilde{\epsilon}_{A_{1}\ldots A_{D}}\tilde{\epsilon}^{B_{1}\ldots B_{D}} \delta^{A_{1}}_{B_{1}}\ldots \delta^{A_{d}}_{B_{d}}\times\, \left [\omega^{\mu_{1}}{}_{B_{d+1}}\Omega_{\mu_{1}}{}^{A_{d+1}}\right ]\ldots\left [\omega^{\mu_{D-d}}{}_{B_{D}}\Omega_{\mu_{D-d}}{}^{A_{D}}\right ]\label{wedgecalculation}
\end{align}
where $0\le d\le D$ and  $A,\mu\in\{1,\ldots,D\}$.}
\begin{align}
S^{(1)}_{\rm mix}&=\int \rd ^{4}x\,\det  (E )\, \tilde{\epsilon}_{abcd_{1}}\tilde{\epsilon}^{abcd_{2}}E^{\mu }{}_{d_{2}}\bar{E}_{\mu }{}^{d_{1}}, \nn
S^{(2)}_{\rm mix}&=\int \rd ^{4}x\,\det  (E)\,\tilde{\epsilon}_{abc_{1}d_{1}}\tilde{\epsilon}^{abc_{2}d_{2}} E^{\mu}{}_{d_{2}}\bar E_{\mu}{}^{d_{1}}\, E^{\nu}{}_{c_{2}}\bar E_{\nu}{}^{c_{1}}, \nn
S^{(3)}_{\rm mix}&=\int \rd ^{4}x\,\det  (E)\,\tilde{\epsilon}_{ab_{1}c_{1}d_{1}}\tilde{\epsilon}^{ab_{2}c_{2}d_{2}} E^{\mu}{}_{d_{2}} \bar E_{\mu}{}^{d_{1}}\,   E^{\nu}{}_{c_{2}}\bar E_{\nu}{}^{c_{1}}\, E^{\rho}{}_{b_{2}}\bar E_{\rho}{}^{b_{1}},
\end{align}
and using the general form of the induced vielbeins \eqref{generalvielbeinform}, the actions reduce to 
\begin{align}
S^{(1)}_{\rm mix}&=\int \rd ^{4}x\,\det  (E)\, \tilde{\epsilon}_{abcd_{1}}\tilde{\epsilon}^{abcd_{2}}\Big[\Phi_{d_{2}}{}^{d_{1}}+\kappa E^{\mu}{}_{d_{2}}\Pi^{d_{1}}\partial_{\mu}\pi\Big],\nn
S^{(2)}_{\rm mix}&=\int \rd ^{4}x\,\det  (E)\,\tilde{\epsilon}_{abc_{1}d_{1}}\tilde{\epsilon}^{abc_{2}d_{2}} \Big [\Phi_{d_{2}}{}^{d_{1}}\Phi_{c_{2}}{}^{c_{1}}+
 2\kappa\Phi_{d_{2}}{}^{d_{1}}E^{\mu}{}_{c_{2}}\Pi^{c_{1}}\partial_{\mu}\pi\Big],\nn
S^{(3)}_{\rm mix}&=\int \rd ^{4}x\,\det  (E)\,\tilde{\epsilon}_{ab_{1}c_{1}d_{1}}\tilde{\epsilon}^{ab_{2}c_{2}d_{2}} \Big[\Phi_{d_{2}}{}^{d_{1}}\Phi_{c_{2}}{}^{c_{1}}\Phi_{b_{2}}{}^{b_{1}}+ 3\kappa\Phi_{d_{2}}{}^{d_{1}}\Phi_{c_{2}}{}^{c_{1}}E^{\mu}{}_{b_{2}}\Pi^{b_{3}}\partial_{\mu}\pi\Big]\ ,
\end{align}
where we have defined $\Phi_{a}{}^{b}\equiv E^{\mu}{}_{a}F_{\nu}{}^{b}(X^{\sigma},\pi)\partial_{\mu}X^{\nu}$, for brevity.  
Performing the Levi-Civita contractions and using brackets to denote traces of $\Phi_{a}{}^{b}$ ($\left [\Phi\right ]\equiv \Phi_{a}{}^{a}$, $\left[\Phi^{2}\right ]\equiv \Phi_{a}{}^{b}\Phi_{b}{}^{a}$, etc.) these can be expressed as
\begin{align}
S^{(1)}_{\rm mix}&=3!\int \rd ^{4}x\,\det  (E)\Big[\left [\Phi\right ]+\kappa E^{\mu}{}_{a}\Pi^{a}\partial_{\mu}\pi\Big],\nn
S^{(2)}_{\rm mix}&=2!\int \rd ^{4}x\,\det  (E) \Big [\left [\Phi\right ]^{2}-\left [\Phi^{2}\right ]+ 2\kappa\left [\Phi\right ]E^{\mu}{}_{a}\Pi^{a}\partial_{\mu}\pi-2\kappa\Phi_{a}{}^{b}E^{\mu}{}_{b}\Pi^{a}\partial_{\mu}\pi\Big],\nn
S^{(3)}_{\rm mix}&=\int \rd ^{4}x\,\det  (E ) \Big[\left [\Phi\right ]^{3}-3\left [\Phi\right ]\left [\Phi^{2}\right ]+2\left [\Phi^{3}\right ]+  3\kappa\left (\left [\Phi\right ]^{2}-\left [\Phi^{2}\right ]\right )E^{\mu}{}_{a}\Pi^{a}\partial_{\mu}\pi\nn
&\quad- 6\kappa\left [\Phi\right ]\Phi_{a}{}^{b}E^{\mu}{}_{b}\Pi^{a}\partial_{\mu}\pi+6\kappa\Phi_{a}{}^{b}\Phi_{b}{}^{c}E^{\mu}{}_{c}\Pi^{a}\partial_{\mu}\pi\Big]\ .
\end{align}

\subsection{Maximally Symmetric Cases}

In this section we specialize to the case of maximally symmetric bulks.  Since every isometry of the bulk metric $G_{AB}$ translates into a symmetry of the galileon field, these are the cases with the highest number of galileon symmetries.  

Using Gaussian normal coordinates and the same conventions as in \eqref{gaussiannormal5Dmetric},  a 5D maximally symmetric metric can always be put in the form
\begin{align}
G=\rd\rho^{2}+f^{2}(\rho)\tilde{g}_{\mu\nu}(X^{\sigma})\rd X^{\mu}\rd X^{\nu}\label{maximallysymmetricfactoredmetric}
\end{align}
where $\tilde{g}_{\mu\nu}(X^{\sigma})$ is a maximally symmetric 4D metric which is independent of $\rho$.  Each surface of constant $\rho$ defines an embedding of a 4D maximally symmetric hypersurface in the bulk.  The various possibilities are enumerated in Figure 2 of \cite{Goon:2011qf}, where more details of the coordinate systems and embeddings are given. 

The 5D vielbeins \eqref{5Dgeneralconstructionmetricandvielbeins} are then given by
\begin{align}
F_{5}{}^{5}=1, \quad
F_{\mu}{}^{5}=0, \quad {\rm and}\quad 
F_{\mu}{}^{a}=f(\rho)f_{\mu}{}^{a}(X^{\sigma})\ ,
\end{align}
where $f_{\mu}{}^{a}(X^{\sigma})$ is defined through $\tilde{g}_{\mu\nu}=f_{\mu }{}^{a}f_{\nu}{}^{b}\eta_{ab}$.  The pullback of the metric as defined in \eqref{pullbackvielbeins} has vielbein components given by 
\begin{align}
\bar{E}_{\mu}{}^{a}&=f(\pi)f_{\nu}{}^{a}(X^{\sigma})\partial_{\mu}X^{\nu}+\kappa(X^{\sigma},\pi)\Pi^{a}\partial_{\mu}\pi  \ ,
\end{align}
with
\begin{align}
\Pi^{a}&=\frac{1}{f(\pi)}f^{\nu a}\frac{\partial x^{\mu }}{\partial X^{\nu}}\partial_{\mu}\pi\nn
\kappa&\equiv \frac{1}{\Pi^{2}}\left [-1+\sqrt{1+\Pi^{2}}\right ]\nn
\Pi^{2}&\equiv \eta_{ab}\Pi^{a}\Pi^{b}\nn
&=\frac{1}{f(\pi)^{2}}\tilde{g}^{\alpha\beta}\frac{\partial x^{\mu }}{\partial X^{\alpha}}\frac{\partial x^{\nu }}{\partial X^{\beta}}\partial_{\mu }\pi\partial_{\nu}\pi\ ,\label{inducedvielbeincomponentsmaxsymstuck}
\end{align}
corresponding to an induced metric of the form
\begin{align}
\bar{g}_{\mu\nu}&=f(\pi)^2\tilde{g}_{\alpha\beta}(X^{\sigma})\partial_{\mu} X^{\alpha}\partial_{\nu} X^{\beta}+\partial_{\mu}\pi\partial_{\nu}\pi\ .
\end{align}
Making these substitutions, the mixing actions of the previous section with all $\pi$ dependence explicitly displayed become
\begin{align} S^{(1)}_{\rm mix}&=3!\int \rd ^{4}x\,\det  (E )\Big[f(\pi)\left [\Psi\right ]+\frac{1}{f(\pi)}\kappa E^{\mu}{}_{a}f^{\alpha a}\frac{\partial x^{\nu}}{\partial X^{\alpha}}\partial_{\nu}\pi \partial_{\mu}\pi\Big],\nn
S^{(2)}_{\rm mix}&=2!\int \rd ^{4}x\,\det  (E ) \Big [f(\pi)^{2}\left (\left [\Psi\right ]^{2}-\left [\Psi^{2}\right ]\right )+ 2\kappa \left [\Psi\right ]E^{\mu}{}_{a}f^{\alpha a}\frac{\partial x^{\nu}}{\partial X^{\alpha}}\partial_{\nu}\pi \partial_{\mu}\pi\nn
&\quad -2\kappa \Psi_{a}{}^{b}E^{\mu}{}_{b}f^{\alpha a}\frac{\partial x^{\nu}}{\partial X^{\alpha}}\partial_{\nu}\pi \partial_{\mu}\pi\Big],\nn
S^{(3)}_{\rm mix}&=\int \rd ^{4}x\,\det  (E ) \Big[f(\pi)^{3}\left (\left [\Psi\right ]^{3}-3 \left [\Psi\right ]\left [\Psi^{2}\right ]+2 \left [\Psi^{3}\right ] \right )\nn
&\quad+ 3\kappa f(\pi)\left (\left [\Psi\right ]^{2}-\left [\Psi^{2}\right ]\right )E^{\mu}{}_{a}f^{\alpha a}\frac{\partial x^{\nu}}{\partial X^{\alpha}}\partial_{\nu}\pi \partial_{\mu}\pi- 6\kappa f(\pi)\left [\Psi\right ]\Psi_{a}{}^{b}E^{\mu}{}_{b}f^{\alpha a}\frac{\partial x^{\nu}}{\partial X^{\alpha }}\partial_{\nu}\pi \partial_{\mu}\pi\nn
&\quad +6\kappa f(\pi)\Psi_{a}{}^{b}\Psi_{b}{}^{c}E^{\mu}{}_{c}f^{\alpha a}\frac{\partial x^{\nu}}{\partial X^{\alpha}}\partial_{\nu}\pi \partial_{\mu}\pi\Big] \ ,\label{maxsymmactions}
\end{align}
where $\Psi_{a}{}^{b}\equiv E^{\mu}{}_{a}f_{\nu}{}^{b}\partial_{\mu}X^{\nu}$ and as before brackets denote traces of $\Psi_{a}{}^{b}$.  

Though complicated, the above actions are useful as they explicitly demonstrate where the St\"uckelberg fields would arise in the procedure of \cite{ArkaniHamed:2002sp} for restoring general coordinate invariance.  The embedding functions $X^{A}$ are the St\"uckelberg fields.  The expressions simplify in the next section where we go to unitary gauge.  

\subsection{Maximally Symmetric Actions in Unitary Gauge  }

Finally, we present explicit expressions for the maximally symmetric mixing actions in unitary gauge \eqref{galileonembeddingfunctions} where the first four embedding functions coincide with the coordinates on $\mathcal{M}$, $X^\mu=x^\mu$.

The induced vielbein components \eqref{inducedvielbeincomponentsmaxsymstuck} in unitary gauge become
\begin{align}
\bar{E}_{\mu}{}^{a}&=f(\pi)f_{\mu}{}^{a}(x^{})+\kappa(\pi,x^{})\Pi^{a}\partial_{\mu}\pi \ ,
\end{align}
where
\begin{align}
\Pi^{a}&=\frac{1}{f(\pi)}f^{\mu a}\partial_{\mu}\pi,\nn
\kappa&\equiv \frac{1}{\Pi^{2}}\left [-1\pm\sqrt{1+\Pi^{2}}\right ],\nn
\Pi^{2}&\equiv \eta_{ab}\Pi^{a}\Pi^{b}=\frac{1}{f(\pi)^{2}}\tilde{g}^{\mu\nu}(x^{}) \partial_{\mu }\pi\partial_{\nu}\pi =\frac{(\partial\pi)^{2}}{f(\pi)^{2}}\ .\label{maxsymmunitarygaugevielbeins}
\end{align}
The actions in \eqref{maxsymmactions} become
\begin{align} S^{(1)}_{\rm mix}&=3!\int \rd ^{4}x\,\det  (E)\Big[f(\pi)\left [\Psi\right ]+\frac{1}{f(\pi)}\kappa E^{\mu}{}_{a}f^{\nu a} \partial_{\nu}\pi \partial_{\mu}\pi\Big],\nn
S^{(2)}_{\rm mix}&=2!\int \rd ^{4}x\,\det  (E ) \Big [f(\pi)^{2}\left (\left [\Psi\right ]^{2}- \left [\Psi^{2}\right ]\right )+ 2\kappa \left [\Psi\right ]E^{\mu}{}_{a}f^{\nu a} \partial_{\nu}\pi \partial_{\mu}\pi\nn
&\quad -2\kappa \Psi_{a}{}^{b}E^{\mu}{}_{b}f^{\nu a} \partial_{\nu}\pi \partial_{\mu}\pi\Big],\nn
S^{(3)}_{\rm mix}&=\int \rd ^{4}x\,\det  (E ) \Big[f(\pi)^{3}\left (\left [\Psi\right ]^{3}-3 \left [\Psi\right ]\left [\Psi^{2}\right ]+2 \left [\Psi^{3}\right ]\right ) \nn
&\quad +3\kappa f(\pi)\left (\left [\Psi\right ]^{2}-\left [\Psi^{2}\right ]\right )E^{\mu}{}_{a}f^{\nu a} \partial_{\nu}\pi \partial_{\mu}\pi- 6\kappa f(\pi)\left [\Psi\right ]\Psi_{a}{}^{b}E^{\mu}{}_{b}f^{\nu a}\partial_{\nu}\pi \partial_{\mu}\pi\nn
&\quad +6\kappa f(\pi)\Psi_{a}{}^{b}\Psi_{b}{}^{c}E^{\mu}{}_{c}f^{\nu a} \partial_{\nu}\pi \partial_{\mu}\pi\Big] \ , 
\label{maxsymmactionsunitarygauge}
\end{align}
where $\Psi_{a}{}^{b}\equiv E^{\mu}{}_{a}f_{\mu}{}^{b}$ and as before brackets denote traces of $\Psi_{a}{}^{b}$.

 \section{Maximally Symmetric Solutions and Fluctuations}\label{maximallysymmetricsolutions}
 
 In this section we examine the maximally symmetric solutions of the full theory and study the fluctuations around these solutions.  In particular, we look for solutions for which the bulk is maximally symmetric and:
\begin{enumerate}
\item the physical vielbein is in a configuration $\bE^{a}=\bE^{a}_{0}$ such that the metric $g_{0}=\bE^{a}_{0}\otimes\bE_{0}^{b}\eta_{ab}$ is maximally symmetric,
\item the galileon field is in a constant configuration $\pi=\pi_0$, so that $\bar{\bE}^{a}_{0}\big|_{\pi=\pi_0}=\Delta\, \bE^{a}$ where $\Delta\equiv f(\pi_0)$ is a constant factor. 
\end{enumerate}
We will see that a massive graviton and a scalar propagate around each of these vacua.
 
 \subsection{Maximally Symmetric Solutions}

We start with the general action \eqref{galileonmetricmixingaction} and, for simplicity, restrict ourselves to cases where $S_{\rm gal}$ is set to zero,\begin{align}
S&=\frac{M_{\rm pl}^{2}}{2}\Big [\int \rd ^{4}x\, \left (\det E_{}{}^{}\right )\,\left( R[E_{}{}^{}]-2\Lambda\right)\nn
&\quad-\frac{m^{2}}{4}\sum_{n=1}^{3}\frac{\beta_{n}}{n!(4-n)!}\int\, \tilde{\epsilon}_{a_{1}\ldots a_{4}}\bar\bE^{a_{1}}\wedge\ldots\wedge\bar\bE^{a_{n}}\wedge\bE^{a_{n+1}}\wedge \ldots\bE^{a_{4}}\Big ]\ .
\end{align}
We work in unitary gauge.  
The equation of motion for the physical vielbein is
\begin{align}
&\frac{M_{\rm pl}^{2}}{2}\Big[2\det(E_{}{}^{})\left (R_{\mu\beta}E^{\beta a}-\frac{1}{2}E_{\mu}{}^{a}R+E_{\mu}{}^{a}\Lambda \right ) \nn
&\quad +\frac{m^{2}}{4}E_{\mu}{}^{a'}E_{\mu'}{}^{a}\tilde{\epsilon}^{\mu'\nu\rho\sigma}\tilde{\epsilon}_{a'bcd}\Big(\frac{\beta_{1}}{2}\bar{E}_{\nu}{}^{b}E_{\rho}{}^{c}E_{\sigma}{}^{d}+\frac{\beta_{2}}{2}\bar{E}_{\nu}{}^{b}\bar E_{\rho}{}^{c}E_{\sigma}{}^{d}+\frac{\beta_{3}}{3!}\bar{E}_{\nu}{}^{b}\bar E_{\rho}{}^{c}\bar E_{\sigma}{}^{d}\Big)\Big]=0,
\end{align}
which after substituting $\bar E_{0\mu}{}^{a}=\Delta E_{0\mu}{}^{a}$ and using the properties of the Ricci tensor for maximally symmetric spaces yields the condition
\begin{align}
\frac{M_{\rm pl}^{2}}{2}\det(E_{0}{}^{}) E_{0\mu}{}^{a}\Big[   -\frac{1}{2} R+2\Lambda   +\frac{m^{2}}{4} \Big(3\beta_{1}\Delta+3\beta_{2}\Delta^{2} +\beta_{3}\Delta^{3} \Big)\Big] &=0\ .\label{examplevielbeineom}
\end{align}
Because the vierbein is invertible, the quantity in square brackets must vanish.

Next, we need to ensure that the $\pi$ equations of motion are satisfied on our desired configuration.  The unitary gauge induced vielbein takes the form
\begin{align}
\bar{E}_{\mu}{}^{a}&=f(\pi)E_{0\mu}{}^{a}(x^{})+\frac{\kappa(\pi)}{f(\pi)}E_{0}^{\nu a}\partial_{\nu}\pi\partial_{\mu}\pi\label{exampleinducedvielbein}\ ,
\end{align}
as in \eqref{maxsymmunitarygaugevielbeins},  and we identify $\Delta=f(\pi_0)$.  Only the mixing action contains the $\pi$ degrees of freedom,
\begin{align}
S_{\rm mixing}&=-\frac{m^{2}M_{\rm pl^{2}}}{8}\sum_{n=1}^{3}\frac{\beta_{n}}{n!(4-n)!}\int\, \tilde{\epsilon}_{a_{1}\ldots a_{4}}\bar\bE^{a_{1}}\wedge\ldots\wedge\bar\bE^{a_{n}}\wedge\bE^{a_{n+1}}\wedge \ldots\bE^{a_{4}}\ ,
\end{align} 
and there are two types of terms appearing in this: those where derivatives act upon $\pi$ and those without any derivatives on $\pi$.  Due to the form of \eqref{exampleinducedvielbein}, any derivative term contains at least two $\pi$'s with a derivative acting upon each field and hence the resulting equations of motion will contain at least one factor of a derivative acting on a field $\pi$.  A constant $\pi$ configuration will therefore automatically solve the equations of motion stemming from these derivative terms.  The non-derivative part of the mixing action, with the physical vielbein evaluated at $\bE_{0}$, takes the form
\begin{align}
S_{\rm mixing}^{\rm non-derivative}&=-\frac{m^{2}M_{\rm pl^{2}}}{8}\sum_{n=1}^{3}\frac{\beta_{n}f(\pi)^{n}}{n!(4-n)!}\int\, \tilde{\epsilon}_{abcd}\bE_{0}^{a}\wedge \bE_{0}^{b}\wedge \bE_{0}^{c}\wedge \bE_{0}^{d} 
\end{align} 
and the $\pi$ equation of motion yields the condition
\begin{align}
&\sum_{n=1}^{3}\frac{f'(\pi_0)f(\pi_0)^{n-1}\beta_{n}}{(n-1)!(4-n)!}=f'(\pi_0)\left (\frac{\beta_{1}}{3!}+\frac{f(\pi_0)\beta_{2}}{2}+\frac{f(\pi_0)^{2}\beta_{3}}{2}\right )=0\ ,
\end{align}
where $f'(\pi_0)=\partial_{\pi}f(\pi_0)$. 

In summary, the $E_{\mu}{}^{a}=E_{0\mu}{}^{a}$, $\pi=\pi_0$ configuration is a solution when
\begin{align}
-\frac{1}{2} R+2\Lambda   +\frac{m^{2}}{4} \Big(3\beta_{1}\Delta+3\beta_{2}\Delta^{2} +\beta_{3}\Delta^{3} \Big)&=0,\label{examplesimplifiedvielbeineomcondition}
\end{align}
and either
\begin{equation}
f'(\pi_0) = 0 \ \ \ \ {\rm or} \ \ \ \ \beta_{1}+3\Delta\beta_{2}+3\Delta^{2}\beta_{3}=0\ ,
\label{twopiconstraints}
\end{equation} 
where $\Delta=f(\pi_0)$.
It should be noted that when $S_{\rm gal}$ is non-trivial, there typically still exist maximally symmetric solutions of our desired form.  The additional terms will only affect the $\pi$ equations of motion, causing them to differ by the addition of couplings appearing in $S_{\rm gal}$.

\subsection{Fluctuation Lagrangian}

We now calculate the Lagrangian for quadratic fluctuations about these maximally symmetric solutions.  We have found that for $\pi=\pi_0=constant$ to be a solution it must satisfy one of the two conditions \eqref{twopiconstraints}. However, the second condition turns out to be problematic: when $S_{\rm gal}=0$, the kinetic term for the galileon fluctuations, $\tilde{\pi}$, arises as $\sim(\beta_{1}+3\Delta\beta_{2}+3\Delta^{2}\beta_{3})(\partial\tilde\pi)^{2}$, and hence this second condition leads to a vanishing canonical kinetic term.  Non-trivial choices of $S_{\rm gal}$ could allow this second condition to be satisfied while retaining a canonical kinetic term, but we shall not consider this possibility here, and shall focus instead on those cases for which $f'(\pi_0)=0$.  A brief survey of Figure 2 of \cite{Goon:2011qf} reveals that this condition can only be satisfied when the $\pi_0$ configuration corresponds to a maximally symmetric brane embedded in a 5D version of itself.  That is, the induced metric either comes from embedding $AdS_{4}$ in $AdS_{5}$, $M_{4 }$ in $M_{5}$ or $dS_{4}$ in $dS_{5}$.   These are the three cases which we analyze in detail, showing that a massive graviton and a scalar propagates on each of these vacua.

We now expand the action to quadratic order in fluctuations about any one of these three scenarios in order to check identify the propagating fluctuations and assess their stability.  There is a redundancy between the $\beta_{n}$ parameters and the parameter $m$ which can be removed by imposing  
\begin{align}
\Delta\beta_{1}+2\Delta^{2}\beta_{2}+\Delta^{3}\beta_{3}&=8\ .\label{gravitonmassnormalizationcondition}
\end{align}
This condition will ensure that the graviton which propagates on this background has mass $m$. 

Defining the fluctuations of the physical vielbein and galileon by 
\be \bE^{a}=\bE_{0}^{a}+\bH^{a}, \ \ \  \pi=\pi_0 +\tilde{\pi},\ee
we expand out to $\mathcal{O}(\tilde{\pi}^{2})$, $\mathcal{O}(\bH\tilde{\pi})$ and $\mathcal{O}(\bH^{2})$ and disregard all cubic and higher terms.  The result from expanding the mixing term is
 \begin{align}
S_{\rm mixing}&=-\frac{M_{\rm pl}^{2}m^{2}}{8}\int\tilde{\epsilon}_{abcd}\Big[\left (  -\frac{\beta_{2} \Delta ^2}{12}+\frac{4}{3}-\frac{1}{4} \beta_{2} \tilde{\pi} ^2 f'' \Delta
   -\frac{1}{6} \beta_{1} \tilde{\pi} ^2 f''+\frac{2 \tilde{\pi} ^2
   f''}{\Delta }\right )\bE^{a}_{0}\wedge\bE^{b}_{0}\wedge\bE^{c}_{0}\wedge\bE^{d}_{0}\nn
   &\quad + \left ( -\frac{\beta_{1}}{3}-\frac{\beta_{2} \Delta }{2}  +\frac{4}{\Delta }\right )\bE^{a}_{0}\wedge\bE^{b}_{0}\wedge\bE^{c}_{0}\wedge\left (\kappa\Pi^{d}\rd\tilde{\pi}\right ) +  \left (\frac{\beta_{2} \Delta ^2}{6}+\frac{\beta_{1} \Delta
   }{3}+\frac{4}{3} \right )\bE^{a}_{0}\wedge\bE^{b}_{0}\wedge\bE^{c}_{0}\wedge\bH^{d}\nn 
   &\quad+  \left (\frac{\beta_{2} \Delta ^2}{4}+\frac{\beta_{1} \Delta }{2}\right ) \bE^{a}_{0}\wedge\bE^{b}_{0}\wedge\bH^{c}\wedge\bH^{d}\Big]\ ,
\end{align}
where \eqref{gravitonmassnormalizationcondition} has been used to eliminate $\beta_{3}$. 

The Einstein-Hilbert and cosmological constant terms are expanded similarly.  We will call $S^{(2)}_{EH}$ the standard quadratic action one would get from expanding Einstein-Hilbert plus a cosmological constant (i.e. the massless graviton action), written in terms of vielbeins, whose explicit form we will not need.
The total quadratic action reads
 \begin{align}
S^{(2)}&=S^{(2)}_{EH}+\frac{M_{\rm pl}^{2}}{2}\int\rd^{4}x\, \det (E_{0}{}^{})\,\Big[\nn
&\quad +m^{2}\left (\delta^{c }_{a}\delta^{d}_{b}-\delta^{d}_{a}\delta^{c}_{b}\right ) E^{\mu}_{0}{}_{c}H_{\mu}{}^{a}E^{\nu}_{0}{}_{d}H_{\nu}{}^{b} -6m^{2}\omega\tilde{\pi}^{2}f''  - \frac{3m^{2}\omega}{2\Delta}g_{0}^{\mu\nu}\partial_{\mu}\tilde{\pi}\partial_{\nu}\tilde{\pi} \Big]\ ,\label{simplifiedquadraticaction}
\end{align} 
We have used the condition \eqref{examplesimplifiedvielbeineomcondition} to eliminate $\beta_{2}$, and
in addition we have used $\kappa=\frac{1}{2}+\mathcal{O}(\tilde\pi^{2})$ and we have defined
\begin{align}
\omega&=  -\frac{R}{2m^2 \Delta }+\frac{2
   \Lambda }{m^2 \Delta }+\frac{4}{\Delta }+\frac{ \beta_{1}}{3}  \ .\label{omegadefinition}
\end{align}
Since we have expanded about a solution, all tadpole terms cancel. 

We have decoupled scalar and metric perturbations.  For the scalar not to be a ghost, we must ensure that 
\be \omega/\Delta>0.\ee
The canonically normalized action becomes
\begin{align}
S^{(2)}  &=S^{(2)}_{EH}+\int\rd^{4}x\, \det (E_{0}{}^{})\,\Big[\nn
&\quad +2m^{2}\left (\delta^{c }_{a}\delta^{d}_{b}-\delta^{d}_{a}\delta^{c}_{b}\right ) E^{\mu}_{0}{}_{c}\hat{H}_{\mu}{}^{a}E^{\nu}_{0}{}_{d}\hat{H}_{\nu}{}^{b}-\frac{1}{2}g_{0}^{\mu\nu}\partial_{\mu}\hat{\pi}\partial_{\nu}\hat{\pi} -2\Delta f''\hat{\pi}^{2}\Big]\ .
\end{align} 
where $\hat{H}_{\mu}{}^{a}\equiv \frac{1}{2}M_{\rm pl} H_{\mu}{}^{a}$ and $\hat{\pi}\equiv \tilde{\pi}M_{\rm pl} m\sqrt{\frac{3\omega }{2\Delta}}$.

The vierbein has sixteen components whereas the metric only has ten, and we would like to eliminate the extra vierbein components. 
The usual metric perturbation $g_{\mu\nu}=g_{0\mu\nu}+2\hat{h}_{\mu\nu}/M_{\rm pl}$ and vierbein perturbation are related by
 \begin{align}
 g_{\mu\nu}&=g_{0\mu\nu}+2\hat{h}_{\mu\nu}/M_{\rm pl}\nn
 &=\eta_{ab}\left (E_{0\mu}{}^{a}+2\hat H_{\mu}{}^{a}/M_{\rm pl}\right )\left (E_{0\nu}{}^{b}+2\hat H_{\nu}{}^{b}/M_{\rm pl}\right )
 \end{align}
 so that $\hat h_{\mu\nu}=2E_{0(\mu}{}^{a}\hat H_{\nu)a}+\mathcal{O}(\hat H^{2})$.  It is convenient to then define $\hat H_{\mu\nu}\equiv \hat H_{\mu a}E_{0\nu}{}^{a}$ so that $\hat{h}_{\mu\nu}=2\hat{H}_{(\mu\nu)}$, i.e. the metric perturbation is the symmetric part of the vierbein perturbation.   In terms of $\hat{H}_{\mu\nu}$, the six antisymmetric components, $a_{\mu\nu}\equiv\hat{H}_{[\mu\nu]}$, are the ones we would like to eliminate. 
 
 The massless graviton action $\mathcal{L}_{EH}^{(2)}$ does not depend on $a_{\mu\nu}$ because it is invariant under linearized local Lorentz transformations which act as a shift on $a_{\mu\nu}$.
 The graviton mass term breaks local Lorentz, and we find
 \begin{align}
 \left (\delta^{c }_{a}\delta^{d}_{b}-\delta^{d}_{a}\delta^{c}_{b}\right ) E^{\mu}_{0}{}_{c}\hat{H}_{\mu}{}^{a}E^{\nu}_{0}{}_{d}\hat{H}_{\nu}{}^{b}=\frac{1}{4}(\hat{h}_{\mu}^{\ \mu})^{2}-\frac{1}{4}\hat{h}_{\mu\nu}\hat{h}^{\mu\nu}+a_{\mu\nu}a^{\mu\nu}\ .
 \end{align}
 We see that the antisymmetric combination $a_{\mu\nu}$ appears as an auxiliary field whose equation of motion sets $a_{\mu\nu}=0$.  The remaining part of the gravitational action is precisely the Fierz-Pauli Lagrangian \cite{Fierz:1939ix} for a massive graviton propagating on a maximally symmetric spacetime.  
\begin{align}
S^{(2)}  &=S^{(2)}_{EH}+\int\rd^{4}x\, \det (E_{0}{}^{})\,\Big[ -\frac{m^{2}}{2}\left (\hat{h}^{\mu\nu}\hat{h}_{\mu\nu}-(\hat{h}_{\mu}{}^{\mu})^{2}\right )  -\frac{1}{2}g_{0}^{\mu\nu}\partial_{\mu}\hat{\pi}\partial_{\nu}\hat{\pi} -2\Delta f''\hat{\pi}^{2}\Big]\ .\label{canonicalsimplifiedquadraticaction}
\end{align}  

The mass of the scalar depends on which of the three maximally symmetric cases we are in.  We look at each in turn: 

\subsubsection{Flat space: $M_{4}$ in $M_{5}$}
The bulk metric for $M_{5}$ is simply
\begin{align}
\rd s^{2}&=\rd\rho^{2}+\eta_{\mu\nu}\rd x^{\mu}\rd x^{\nu}
\end{align}
 so that $f(\pi)=1$ and
\begin{align}
\bar{E}_{\mu}{}^{a}&=\delta_{\mu}^{a}+\kappa\eta^{\nu a}\partial_{\nu}\pi\partial_{\mu}\pi \ .
\end{align} 

Since $\Delta=1$ and $R=0$ we find
\begin{align}
\omega&=\frac{2\Lambda}{m^{2}}+4+\frac{\beta_{1}}{3}
\end{align}
and choosing $\beta_{1}$ so that $\omega> 0$ the galileon sector is healthy and the total canonically normalized quadratic action \eqref{canonicalsimplifiedquadraticaction} becomes
\begin{align}
S^{(2)}  &=S^{(2)}_{EH}+\int\rd^{4}x\, \,\Big[-\frac{m^{2}}{2}\left (\hat{h}^{\mu\nu}\hat{h}_{\mu\nu}-(\hat{h}_{\mu}{}^{\mu})^{2}\right )  -\frac{1}{2}\eta^{\mu\nu}\partial_{\mu}\hat{\pi} \partial_{\nu}\hat{\pi}\Big]\ .
\end{align}  
 This is a massive graviton of mass $m$ and a free decoupled massless scalar.

\subsubsection{Positive Curvature: $dS_{4}$ in $dS_{5}$}

The bulk metric for $AdS_{5}$ can be written
\begin{align}
\rd s^{2}=  \rd\rho^{2}+\left (\frac{\mathcal{R}}{L}\right )^{2}\sin^{2}\left (\frac{\rho}{\mathcal{R}}\right )\left [L^{2}\rd s^{2}_{dS_{4}}\right ] ,
\end{align}
where $\rho\in(0,\pi\mathcal{R})$.  $\mathcal{R}$ is the bulk curvature radius and $L^{2}\rd s^{2}_{dS_{4}}$ is a 4D de Sitter metric with curvature radius $L$ and Ricci curvature $R=12/L^{2}$.  In this case, $f(\pi)=\frac{\mathcal{R}}{L}\sin(\pi/\mathcal{R})$ and we consider a solution where the physical vielbein is in the configuration $E_{0\mu}{}^{a}$ corresponding to the $L^{2}\rd s^{2}_{dS_{4}}$ metric and $\pi$ is expanded about the point $\pi_{0}=\pi\mathcal{R}/2$ so that $f'=0$ and $f''=-1/\left (L\mathcal{R}\right )$.

The canonically normalized quadratic action \eqref{canonicalsimplifiedquadraticaction} is then
\begin{align}
S^{(2)}  &=S^{(2)}_{EH}+\int\rd^{4}x\, \det (E_{0}{}^{})\,\Big[ -\frac{m^{2}}{2}\left (\hat{h}^{\mu\nu}\hat{h}_{\mu\nu}-(\hat{h}_{\mu}{}^{\mu})^{2}\right ) -\frac{1}{2}g_{0}^{\mu\nu}\partial_{\mu}\hat{\pi}\partial_{\nu}\hat{\pi} +\frac{2}{L^{2}}\hat{\pi}^{2}\Big]\ .
\end{align}  
and we have chosen parameters such that $\omega>0$ where
\begin{align}
\omega&=-\frac{6}{m^{2}\mathcal{R}L}+\frac{2L\Lambda}{m^{2}\mathcal{R}}+\frac{4L}{\mathcal{R}}+\frac{\beta_{1}}{3}\ .
\end{align}
 This is a massive graviton of mass $m$ and a free decoupled scalar with mass squared $-4/L^{2}$.
Therefore, the quadratic fluctuations about this solution exhibit a tachyonic instability in the galileon sector with time scale $\sim 1/m_{\pi}\sim L$.

\subsubsection{Negative curvature $AdS_{4}$ in $AdS_{5}$}
The bulk metric for $AdS_{5}$ can be written
\begin{align}
\rd s^{2}=\rd\rho^{2}+\left (\frac{\mathcal{R}}{L}\right )^{2}\cosh^{2}\left (\frac{\rho}{\mathcal{R}}\right )\left [L^{2}\rd s^{2}_{AdS_{4}}\right ],
\end{align}
where $\rho\in(-\infty,\infty)$.  $\mathcal{R}$ is the bulk curvature radius and $L^{2}\rd s^{2}_{AdS_{4}}$ is a 4D anti-de Sitter metric with curvature radius $L$ and Ricci curvature $R=-12/L^{2}$.  In this case, $f(\pi)=\frac{\mathcal{R}}{L}\cosh(\pi/\mathcal{R})$ and we consider a solution where the physical vielbein is in the configuration $E_{0\mu}{}^{a}$ corresponding to the $L^{2}\rd s^{2}_{AdS_{4}}$ metric and $\pi$ is expanded about the point $\pi_{0}=0$ so that $f'=0$ and $f''=1/\left (L\mathcal{R}\right )$.

The canonically normalized quadratic action \eqref{canonicalsimplifiedquadraticaction} is then
\begin{align}
S^{(2)}  &=S^{(2)}_{EH}+\int\rd^{4}x\, \det (E_{0}{}^{})\,\Big[ -\frac{m^{2}}{2}\left (\hat{h}^{\mu\nu}\hat{h}_{\mu\nu}-(\hat{h}_{\mu}{}^{\mu})^{2}\right ) -\frac{1}{2}g_{0}^{\mu\nu}\partial_{\mu}\hat{\pi}\partial_{\nu}\hat{\pi} -\frac{2}{L^{2}}\hat{\pi}^{2}\Big]\ ,
\end{align}  
and we have chosen parameters such that $\omega>0$ where
\begin{align}
\omega&=\frac{6}{m^{2}\mathcal{R}L}+\frac{2L\Lambda}{m^{2}\mathcal{R}}+\frac{4L}{\mathcal{R}}+\frac{\beta_{1}}{3}\ .
\end{align}
This is a massive graviton of mass $m$ and a free decoupled scalar with mass squared $4/L^2$.
Therefore, the quadratic fluctuations about this solution are stable

Note that the quadratic actions for the scalar in all three cases are exactly those found in \cite{Goon:2011qf}, and are invariant under the lowest order part of the non-linearly realized symmetries whose explicit form is given there.  Here, the difference is that we now have a massive graviton propagating as well, which is coupled to the galileon non-linearly in a way which preserves the galileon symmetries..

\section{Self-accelerating Cosmological Solutions and Perturbations}\label{selfacceleratingsolutionsandperturbations}

In this final section we ask whether the galileon massive gravity action \eqref{matrixformofgalileongravitonaction} can drive a stable self-accelerated expansion of the universe.  In the case of pure dRGT massive gravity, there exist self-accelerating solutions \cite{deRham:2010tw,Koyama:2011xz,Nieuwenhuizen:2011sq,Chamseddine:2011bu,D'Amico:2011jj,Gumrukcuoglu:2011ew,Berezhiani:2011mt} where the Hubble constant is set by the graviton mass, $H\sim m$.  The full theory has five degrees of freedom, but on these self-accelerating solutions only the transverse-traceless tensor mode of the graviton propagates -- the scalar and vector degrees of freedom have vanishing kinetic terms \cite{Gumrukcuoglu:2011zh,D'Amico:2012pi,DeFelice:2012mx,Wyman:2012iw,Khosravi:2012rk,Fasiello:2012rw}.    The vectors and scalars are classically strongly coupled around these backgrounds.  

It is known that some extensions of dRGT are able to restore these vanishing kinetic terms \cite{Gumrukcuoglu:2013nza,DeFelice:2013tsa,DeFelice:2013dua,Gabadadze:2014kaa}.
Here we ask whether the addition of the galileons can restore the vanishing kinetic terms.  In \cite{Andrews:2013uca} this question was asked for the case of a flat 5D metric, and it was found that the galileon terms cannot restore the vanishing kinetic terms.
Here we ask the same question in a more general manner by allowing the 5D background metric to be of a much more general form -- a gaussian normal metric where the leaves are an arbitrary Friedmann-Robertson-Walker (FRW) metric with unspecified spatial curvature.  

The physical metric will be of the FRW form with the same sign spatial curvature as the background.  After finding the equations of motions and conditions required for
self-acceleration, we derive 
the quadratic action for perturbations
about these solutions and discuss their properties, finding that even in
this more general setup the 
kinetic terms cannot be restored.  In order to more easily make contact with previous work \cite{Gumrukcuoglu:2011ew,DeFelice:2012mx,DeFelice:2013awa}, the analysis of this section is performed in the metric language discussed at the beginning of Section \ref{MassiveGravityandInteractingSpin2s}.

\subsection{Setup}

We start with the action \eqref{matrixformofgalileongravitonaction},
\be S=S_{\rm GR}[{g}]+S_{\rm mix}[g,\bar{g}]+S_{\rm gal}[\bar{g}],\label{frweaction}\ee
where
\begin{align}
  S_{\rm GR}&=\frac{M_{\rm pl}^{2}}{2}\int \rd ^{4}x\, \sqrt{-g}\left [R[g]-2\Lambda\right ], \\
  S_{\rm mix}&=-\frac{M_{\rm pl}^{2}m^{2}}{8}\int \rd ^{4}x\, \sqrt{-g}\sum_{n=1}^{3}\beta_{n}S_{n}\left (\sqrt{g^{-1}\bar{g}}\right )\ ,\label{matrixformofgalileongravitonaction2} \\
  S_{\rm gal}&=\sum_{i=1}^{5}\int\rd^{4}x\,c_{i}\mathcal{L}_{i}, 
  \end{align}
with the $\mathcal{L}_{i}$ as defined in \eqref{lovelock} and \eqref{galileontadpoleterm}.

The bulk metric will be restricted to take the Gaussian normal form
\begin{align}
G_{AB}\rd X^A\rd X^B = \rd\rho^2+F(\rho)^2f_{\mu\nu}(x)\rd x^\mu \rd x^\nu\ ,
\label{cosmo-5dmetric}
\end{align}
and we  
choose the unitary gauge \eqref{galileonembeddingfunctions} so that the bulk coordinates $X^A$ are related to the coordinates on the 3-brane $x^\mu$ through
$X^\mu(x) = x^\mu$, $X^5(x) = \pi(x)$,
and the induced metric takes the form 
\begin{align}
\bar{g}_{\mu\nu} &=G_{AB} \frac{\partial X^A}{\partial x^\mu}\frac{\partial X^B}{\partial x^\nu} = F(\pi)^2f_{\mu\nu} +\partial_\mu \pi\partial_\nu \pi\ .
\end{align}

We consider the case where the tensor $f_{\mu\nu}$ takes the FRW form, 
\begin{align}
f_{\mu\nu}\rd x^\mu \rd x^\nu = -n(t)^2\rd t^2+\alpha(t)^2 \Omega_{ij}\rd x^i\rd x^j\ ,\label{backfrwans}
\end{align}
where the spatial metric has constant curvature $K$, 
\begin{align}
 \Omega_{ij} \equiv \delta_{ij} + 
  \frac{K\delta_{il}\delta_{jm}x^lx^m}{1-K\delta_{lm}x^lx^m}\ .\label{spatialmetricdefi}
\end{align}

The detailed form of the galileon Lagrangians $S_{\rm gal}$ for the metric \eqref{cosmo-5dmetric} were derived in \cite{Goon:2011qf}.  We will not need them for our argument.  All we will need is the fact that $S_{\rm gal}$ depends only on $\pi$, and contains none of the degrees of freedom in the dynamical metric.

\subsection{Background Cosmology}

We now look for cosmological solutions.  We take our physical metric to be an FRW metric with the same sign spatial curvature as the physical metric
\begin{align}
g_{\mu\nu}\rd x^\mu \rd x^\nu = -N(t)^2\rd t^2+a(t)^2 \Omega_{ij}\rd x^i\rd x^j,\label{physfrw}
\end{align}
where $\Omega_{ij}$ is the spatial metric \eqref{spatialmetricdefi}.
In addition, we assume that the galileon field 
depends only on time, 
\be \pi=\pi(t).\label{pians}\ee

Plugging the ans\"atze \eqref{physfrw}, \eqref{pians} and \eqref{backfrwans} into the action \eqref{frweaction}, we obtain a mini-superspace action (which we do not write here) whose three dynamical variables are the lapse and scale factor of the physical metric and the galileon field, $N$, $a$ and  $\pi$, respectively.  The lapse and scale factor of the background metric, $n$ and $\alpha$, respectively, also appear in the action but are non-dynamical.  There is no time-reparametrization invariance (i.e. we have not introduced St\"uckelberg fields here).

It is convenient to introduce the following quantities
\begin{align}
&\quad X \equiv \frac{\alpha\ F}{a}\ ,\qquad
r\equiv \frac{a\ \tilde{n}}{\alpha\ F\ N}\ ,\qquad
\tilde{n}\equiv \sqrt{F^2\ n^2-\dot{\pi}^2}\ ,\qquad
H \equiv \frac{\dot{a}}{a\ N} \ ,\qquad
H_f \equiv \frac{\dot{\alpha}}{\alpha\ \tilde{n}} \ ,\qquad
\nn
&\quad\rho_g \equiv \frac{X}{8}\left(3\ \beta_1 +3\ \beta_2\ X+\beta_3\ X^2\right)\ , \qquad \qquad J_\phi \equiv \frac{1}{8}\left(\beta_1 +2\ \beta_2\ X+\beta_3\ X^2\right)\ .\label{variousquantitiesforselfacceleration} 
\end{align}

Varying the mini-superspace action with respect to the lapse function $N$ yields a Friedmann equation,
\begin{align}
3\ \left(H^2+ \frac{K}{a^2}\right)=\Lambda+m^2\rho_g\ ,
\label{eqn}
\end{align}
while varying with respect to the scale factor $a$ and then combining with the above equation gives an acceleration equation,
\begin{align}
2\ \left(\frac{\dot{H}}{N}-\frac{K}{a^2}\right) = m^2 J_\phi X\ (1-r)\ .
\label{eqa}\end{align}
We note that the background equations (\ref{eqn}) and (\ref{eqa}) are identical to their counterparts in pure dRGT, except that the definitions of $X$ and $r$ are different \cite{Gumrukcuoglu:2011ew}. The scalar field $\pi$ is determined by the $\pi$ equation of motion (which 
includes only up to second time derivatives due to the ghost-free structure of the galileon terms), which we will not need explicitly.

By combining \eqref{eqa} with the derivative of \eqref{eqn}, we obtain the following constraint equation
\begin{align}
J_\phi\ \left(H_fX-H+\frac{F'\ \dot{\pi}}{F\ N\ r}\right)=0\ ,
\label{eqstuckelberg}
\end{align}
which defines two branches of solutions according to whether $J_\phi=0$ or the quantity in parenthesis is zero.  The definition of $J_\phi$ \eqref{variousquantitiesforselfacceleration} shows that the quantity $X$ is constrained to be constant in time on the $J_\phi=0$ branch.  As a result, the effective energy density from the interaction term $\rho_g$ \eqref{variousquantitiesforselfacceleration} acts as a cosmological constant, yielding a self-accelerating cosmology in the absence of a 
genuine cosmological constant $\Lambda$ in the Lagrangian.  This is the self-accelerating branch.  In the following, we study the perturbations on top of solutions in this branch.

\subsection{Perturbations}

We now introduce perturbations to the self-accelerating background discussed above.  We denote by $\pi$ the background value of the scalar field and $\delta\pi$ the perturbation.   The perturbations to the $00$, $0i$ and $ij$ components of the physical metric will be captured by the fields $\Phi$, $V_i$ and $H_{ij}$ respectively.  We write the perturbed metrics as 
\begin{align}
 g_{\mu\nu}\rd x^{\mu}\rd x^{\nu} 
  & =  -N^2  (1+2\Phi)  \rd t^2
  + 2N  a  V_i \rd t\rd x^i
  + a^2 (\Omega_{ij}+H_{ij}) \rd x^i\rd x^j\ , \nonumber\\
\bar{g}_{\mu\nu}\rd x^{\mu}\rd x^{\nu} 
  & =  F^2(\pi+\delta\pi)\left[-  n^2  \rd t^2
  + \alpha^2 \Omega_{ij}\rd x^i\rd x^j\right] + \partial_\mu (\pi+\delta\pi)\partial_\nu (\pi+\delta\pi)\rd x^\mu \rd x^\nu\ .
\end{align}

For our purposes, it is sufficient to consider only the mixing term between the metrics $g$ and $\bar{g}$, which reads, up to quadratic order in perturbations,
\begin{align}
\frac{S_{\rm mixing}}{M_{\rm pl}^2m^2} &=  \int \rd^4x Na^3\sqrt{\Omega} \left[-\rho_g\ \frac{\left(\sqrt{-g}\right)^{(2)}}{N\ a^3\ \sqrt{\Omega}}
-\rho_f\ \frac{\left(\sqrt{-\bar{g}}\right)^{(2)}}{\tilde{n}\ \alpha^3\ F^3\ \sqrt{\Omega}}+\frac{1}{2}\ X\ J_\phi\Delta\right]\nonumber\\
& \!+\!\frac{1}{8}\! \int \rd^4x N a^3 \sqrt{\Omega} \ M_{GW}^2\ \left[{\rm Tr}[H]^2-H_{ij}H^{ij}-\frac{8\ F'}{F}{\rm Tr}[H]\delta\pi+\frac{24\ F^{\prime\ 2}}{F^2}\ \delta\pi^2\right]\ ,
\label{quadactionmix}\end{align}
where we have defined
\begin{align}
\rho_f &\equiv \frac{r\ X}{8}\left(\beta_1 + 3\ X\ \beta_2+3\ X^2\ \beta_3+X^3\ \beta_4\right)\ ,\nonumber\\
M_{GW}^2 &\equiv \frac{r-1}{8}\ X^2(\beta_2+X\ \beta_3) + X\ J_\phi\ ,
\end{align}
all spatial indices are raised and lowered by $\Omega_{ij}$ and its inverse, and the trace is ${\rm Tr}[H]\equiv\Omega^{ij}H_{ij}$. In \eqref{quadactionmix}, $\left(\sqrt{-g}\right)^{(2)}$ and $\left(\sqrt{-\bar{g}}\right)^{(2)}$ stand for the expansions of the 
square root of determinants up to second order (whose precise expressions are not needed for our 
purposes), and $\Delta$ is a quantity which multiplies $J_\phi$, whose form is not needed because $J_\phi=0$ on the self-accelerating backgrounds we are considering\footnote{For completeness, the expressions are \begin{align}
 \frac{\sqrt{-g}}{N\  a^3\  \sqrt{\Omega}} & = 
  1 + \left(\Phi+\frac{1}{2}{\rm Tr}[H]\right)
  + \left[
     -\frac{1}{2}\Phi^2 + \frac{1}{2}V^iV_i
     +\frac{1}{8}\left({\rm Tr}[H]^2- 2\ H_{ij}H^{ij}\right)
     + \frac{1}{2}\Phi\  {\rm Tr}[H]\right] \ , 
\nonumber\\
 \frac{\sqrt{-\bar{g}}}{\tilde{n}\  \alpha^3\  F^3\ \sqrt{\Omega}} & = 
  1 + \left(\frac{F'}{F}\left(4+\frac{\dot{\pi}^2}{\tilde{n}^2}\right)\delta\pi- \frac{\dot{\pi}}{\tilde{n}^2} \delta{\dot\pi}\right)+ 
\left\{\frac{F^{\prime\ 2}}{2\ F^2}\left(12+5\ \frac{\dot{\pi}^2}{\tilde{n}^2}-\frac{\dot{\pi}^4}{\tilde{n}^4}\right)+\frac{F''}{2\ F}\left(4+\frac{\dot{\pi}^2}{\tilde{n}^2}\right)\right\}\delta\pi^2
\nonumber\\
&\quad-\frac{F'}{F}\ \frac{\dot{\pi}}{\tilde{n}^2}\left(2-\frac{\dot{\pi}^2}{\tilde{n}^2}\right)\delta\pi\ \delta\dot{\pi}
-\left.
\frac{1}{2}\left(1+\frac{\dot{\pi}^2}{\tilde{n}^2}\right)\left(\frac{\delta\dot{\pi}^2}{\tilde{n}^2}-\frac{D_i\delta\pi\ D^i\delta\pi}{a^2\ X^2}\right)
\right.\label{firstmixingperturbations}
																																						     \end{align}
and
\begin{align}
\Delta &\equiv  (1-r)\left({\rm Tr}[H]-\frac{6\ F'}{F}\delta\pi\right)\nonumber\\
&\quad+ \Phi\ {\rm Tr}[H]+\frac{1}{r+1}\ V^iV_i+\frac{1-r}{4}\left({\rm Tr}[H]^2-2\ H_{ij}H^{ij}\right)-\frac{6\ F'}{F}\ \Phi\ \delta\pi
\nonumber\\
&\quad-r\ {\rm Tr}[H]\left[\frac{F'}{F}\left(1+\frac{\dot{\pi}^2}{\tilde{n}^2}\right)\delta\pi-\frac{\dot{\pi}}{\tilde{n}^2}\delta\dot{\pi}\right]
-\frac{2\ r\ \dot{\pi}}{a\ (r+1)\ X\ \tilde{n}}\ V^i\ D_i\delta\pi\nonumber\\
&\quad+\frac{1}{a^2\ X^2(r+1)}\left(r^2-1+r^2\ \frac{\dot{\pi}^2}{\tilde{n}^2}\right)D_i\delta\pi\ D^i\delta\pi-\frac{6\ r\ F'\ \dot{\pi}}{F\ \tilde{n}^2}\ \delta\pi\delta\dot{\pi}\nonumber\\
&\quad+3\ \left[2\ \frac{F^{\prime\ 2}}{F^2}\left(2\ r-1+r\ \frac{\dot{\pi}^2}{\tilde{n}^2}\right)+(r-1)\frac{F''}{F}\right]\delta\pi^2\ .
\end{align}}.

We now argue that this action (plus the Einstein-Hilbert action and galileon action expanded to quadratic order in fluctuations) propagates at most three degrees of freedom: there is always a non-ghost transverse-traceless tensor, and a scalar which may be ghostly, healthy or vanishing depending on the coefficients $c_i$ of the galileon terms.  This is in contrast to the full theory which propagates six degrees of freedom.

To make the argument, first consider what would happen if we were working with cosmological perturbations of pure GR plus cosmological constant.  We break $V_i$ into transverse and longitudinal parts, $V_i=V_i^T+\partial_i V$, and $H_{ij}$ into transverse traceless, longitudinal and trace parts, $H_{ij}=h_{ij}^{TT}+ {1\over 2}(\nabla_i E^T_j+\nabla_j E^T_i)+
2\,\delta_{ij}\,\Psi +\left(\nabla_i\nabla_j - \frac{1}{3}\delta_{ij}\,\nabla^2\right)E$.  In the vector sector, $V_i^T$ would appear with no time derivatives and could be eliminated with its own equations of motion.  

In GR there are no dynamical vector modes, so doing this leaves only the gauge dependent degree of freedom $E_i^T$, resulting in an action consisting only of boundary terms. A similar remark goes through for the scalar modes: $\Phi$ and $V$ appear with no time derivatives and can be eliminated with their own equations of motion, leaving an action depending on the two degrees of freedom $E$ and $\Psi$; these two degrees of freedom correspond to the two gauge degrees of freedom in the scalar sector, and the resulting action quadratic in the scalar modes vanishes up to boundary terms.

Now we come back to our quadratic Lagrangian.  Since $J_\phi=0$ implies $X={\rm constant}$, the first term in \eqref{quadactionmix} corresponds to perturbations of a cosmological constant term, just as it would appear in pure GR with a cosmological constant.  The second term, the perturbations of the fiducial metric determinant, contain only galileon perturbations $\delta\pi$.  The third term $J_\phi\Delta$ vanishes on the self-accelerating background $J_\phi=0$.   The terms in the final line contain no time derivatives, and contain no factors of the lapse or shift $\Phi,V_i$.   The perturbations to the galileon term, which we have not written, contain only $\delta\pi$.  We can see that our quadratic action contains no terms beyond those of GR which depend on the lapse $\Phi$ or the shift $V_i$, thus  
equations of motion for $\Phi$ and $V_i$ will not undergo a modification with respect to GR. As a result, upon integration of these non-dynamical fields, the combination of the first term in \eqref{quadactionmix} and the Einstein-Hilbert term will vanish, up to boundary terms.

After integrating out $\Phi$ and $V_i$, the only dependence on 
the scalar and vector metric perturbations 
is non-derivative, and 
arises from the second line of \eqref{quadactionmix}. Using the equations of motion for these non-dynamical degrees of freedom ($E^T_i$ from vector perturbations, and $\Psi$ and $E$ from scalar perturbations), we are left with the action of the tensor modes with a time dependent mass $M_{GW}$,  
and the action for $\delta\pi$ which consists of the second term of \eqref{quadactionmix} and the galileon terms. \footnote{The inclusion of matter does not change this conclusion. If matter fields minimally coupled to the physical metric are present, the combination of the first term in \eqref{quadactionmix}, the Einstein-Hilbert term and the matter action can be written in terms of the gauge-invariant variables of GR \cite{Gumrukcuoglu:2011zh}.
On the other hand, the second line in Eq.~\eqref{quadactionmix} contains non-derivative contributions to the four gauge-dependent degrees of freedom and the tensor perturbations. In that case, the dynamical degrees of freedom are the two (massive) tensor polarizations of the graviton, the galileon and the matter degrees.}

This is exactly the conclusion in the self-accelerating branch of dRGT theory \cite{Gumrukcuoglu:2011ew,Gumrukcuoglu:2011zh}. Therefore, we expect that one of the missing degrees of freedom in the linearized setup to exhibit an instability at non-linear order \cite{DeFelice:2012mx,DeFelice:2013awa}.

\section{Conclusions}\label{Conclusionslabel}

In order to couple galileons or DBI scalars to the metric in a manner which preserves galileon symmetries and is ghost free, it appears necessary that the graviton be massive~\cite{Gabadadze:2012tr}. In this paper we have rephrased the construction of~\cite{Gabadadze:2012tr} by using the interacting vielbein formalism of \cite{Hinterbichler:2012cn}, thereby avoiding the use of unwieldy matrix square roots.  The vielbein variables are naturally suited to describe galileon-graviton interactions and reproduce the results of \cite{Gabadadze:2012tr} while also making calculation and the explicit construction of the action more efficient.  After explicitly calculating the generic action of the fully non-linear theory and examining the global symmetry properties, we have demonstrated the existence of maximally symmetric solutions and have analyzed their perturbations, showing that they propagate a massive graviton and a non-ghost scalar with negative, zero or positive mass squared for de Sitter, flat, and anti-de Sitter background respectively, and with the magnitude of the mass squared of order the background curvature.  Finally, we have found self-accelerating cosmological solutions of the full non-linear theory and examined their perturbations, showing that, like in pure dRGT theory, the vector and scalar modes have vanishing kinetic terms.  The vanishing of kinetic terms around self-accelerating solutions seems to be a generic feature of theories with intact, geometrically interpretable, non-linearly realized symmetries.

{\bf Acknowledgments:}
AEG acknowledges financial support from the European Research Council
under the European Union's Seventh Framework Programme (FP7/2007-2013) /
ERC Grant Agreement n. 306425 ``Challenging General Relativity''. 
The work of AEG and SM was supported by the World Premier International Research Center Initiative (WPI Initiative), MEXT, Japan. 
SM also acknowledges the support by Grant-in-Aid for Scientific Research 24540256 and 21111006. 
The work of MT is supported in part by the US Department of Energy and NASA ATP grant NNX11AI95G.  
Research at Perimeter Institute is supported by the Government of Canada through Industry Canada and by the Province of Ontario through the Ministry of Economic Development and Innovation. This work was made possible in part through the support of a grant from the John Templeton Foundation. The opinions expressed in this publication are those of the authors and do not necessarily reflect the views of the John Templeton Foundation (KH).  
KH and MT would like to thank the Institute for the Physics and Mathematics of the Universe (IPMU) at the University of Tokyo, where this collaboration began, for their wonderful hospitality.
\vspace*{-.1in}

\bibliographystyle{utphys}
\addcontentsline{toc}{section}{References}
\bibliography{mGRgalileons_arxivfinal}

\end{document}